\documentclass[prc,aps,a4paper,groupedaddress,superscriptaddress,nofootinbib,showpacs
,preprintnumbers,twocolumn]{revtex4}
\usepackage{graphicx}
\usepackage{amsfonts}
\usepackage{amssymb}
\usepackage{amsmath}
\usepackage{natbib}
\usepackage{dcolumn}
\usepackage{bm} 
\usepackage{color,xcolor}
\usepackage{multirow}

\newcommand{\bwt}{\begin{widetext}}
	\newcommand{\ewt}{\end{widetext}}
\newcommand{\beq}{\begin{equation}}
	\newcommand{\eeq}{\end{equation}}
\newcommand{\bea}{\begin{eqnarray}}
	\newcommand{\eea}{\end{eqnarray}}

\begin{document}
	
\title{Pasta phases in neutron stars under strong magnetic fields}
\author{Xiaopeng Wang}
\author{Jing Li}
\author{Jianjun Fang}
\email{jian-junfang@163.com}
\affiliation{School of Physics and Physical Engineering, Qufu Normal University, 273165 Qufu, China}
\author{Helena Pais}
\email{hpais@uc.pt}
\author{Constan\c ca Provid{\^e}ncia}
\email{cp@uc.pt}
\affiliation{CFisUC, Department of Physics, University of Coimbra, 3004-516 Coimbra, Portugal}

\begin{abstract}
	 In the present work, we consider nuclear matter in the innermost crust of neutron stars under the presence of a strong magnetic field within the framework of a relativistic mean-field description. 
	 Two models with a different slope of the symmetry energy are considered in order to discuss the density-dependence of the equation of state on the crust structure. The non-homogeneous matter in $\beta$-equilibrium  is described within the coexisting phases method, and the effect of including the anomalous magnetic moment is discussed. Five different geometries for the pasta structures are considered. It is shown that strong magnetic fields cause an extension of the inner crust of the neutron stars, with the occurrence of a series of disconnected non-homogeneous matter regions above the one existing for a null magnetic field. Moreover, we observed that in these disconnected 
	 regions, for some values of the magnetic field, all five different cluster geometrical shapes occur, and the gas density is close to the cluster density. Also, the  pressure at the neutron star crust-core transition much larger than the pressure obtained for a zero magnetic field. Another noticeable effect of the presence of strong magnetic fields is the increase of the proton fraction, favoring the appearance of protons in the gas background.   
\end{abstract}
	\maketitle
	
\section{INTRODUCTION}

In different astrophysical sites, such as core-collapse supernovae \cite{watanabe,pais}, neutron stars (NSs) \cite{lattimer}, or even NS mergers, non-homogeneous  matter in different geometric structures, the so-called pasta phases, may exist under given conditions of temperature, density and asymmetry \cite{ravenhall,cini,grill,okamoto}.  These structures result from the competition of the strong force and the Coulomb interaction. The nuclear shapes evolve  monotonically with increasing density from droplets to rods, slabs, tubes and bubbles. It has been shown that  the presence of  pasta structures may be related with important mechanisms in compact stars, as for instance the decay of the magnetic field \cite{pons2013},  neutrino opacity, thermal and electrical conductivity, superfluidity, see \cite{Chamel} for a review. These mechanisms can give information on the thermal evolution of stars, or the timing and structure of glitches \cite{Chamel,Lattimer:1994glx,Reddy:1997yr,Reddy:1997gd,Arzoumanian:2009qn,Page:2013hxa,Page:2000wt}, thus making the low-density non-homogeneous matter equation of state (EoS) an important issue in the nuclear astrophysics fields.

 Magnetars \cite{duncan,thompson,usov,paczynski}, mainly Soft Gamma Repeaters (SGRs) and Anomalous X-ray Pulsars (AXPs), belong to a kind of NSs with a very strong magnetic field at the surface,  up to   $10^{14} \sim 10^{15}$ G and quite long spin periods, of the order of $2\sim 12 $s \cite{Kaspi2017}. Nowadays  about thirty of such objects have been observed \cite{mcgill}. Frequently, these objects emit large amounts of electromagnetic radiation in the form of x-rays, or $\gamma$-rays. More recently, even a fast radio burst (FRB), i.e a radio wave emission with a duration of circa a millisecond, detected by the Canadian Hydrogen Intensity Mapping Experiment (CHIME) FRB project,  was  associated with the magnetar SGR 1935+2154 \cite{frb}.
 The detection of the gravitational wave GW170817 emitted by a binary neutron star merger \cite{abbott2017}, followed up by the detection of the electromagnetic counterpart, the $\gamma$-ray
burst GRB170817A \cite{grb}, and the electromagnetic
transient AT2017gfo \cite{kilo}, has opened the new era of multi-messenger astrophysics. With the next generation of GW detectors, it is expected that besides transient events also the continuous gravitational wave emission will be detected.
In particular, magnetars are good candidates to be a source of continuous gravitational wave emission, and we expect that in the future it will be possible to detect this type of gravitational waves.
 
The scalar virial theorem \cite{lai,shapiro} provides an upper limit on the central magnetic field strength of NSs, $\sim 10^{18}$ G. Other calculations, like the solutions of the coupled Einstein-Maxwell equations \cite{lorene1,lorene2} confirmed that in the center of a NS, the magnetic field has an intensity not higher than $10^{18}$ G \cite{broderick02,chatterjee15,gomes19}. Presently, it is believed that it is necessary a   mixed poloidal-toroidal configuration for stable magnetized NS  \cite{lander11,uryu19}, the relative magnitude of each component being dependent on the boundary conditions imposed \cite{haskell08,lander09}, and, therefore, on the structure of the NS. In \cite{uryu19} the maximum strength of both components, poloidal and toroidal, do not differ much. However, it was shown that in the presence of a differentially rotating core the toroidal magnetic field can be amplified due to the winding, so that  toroidal fields that dominate over the poloidal ones may be generated \cite{Kotake2006,Uryu2021}. In \cite{Ciolfi2013}, the authors could also  obtain  a larger  toroidal component  than the poloidal one within a  twisted-torus equilibrium configuration for nonrotating magnetized NSs.

Although realistic magnetic fields, such as the ones that may exist inside NS, do not affect much the core EoS, as discussed in \cite{broderick}, it has been shown that these fields have a non-negligible effect, both on the outer and inner  crusts.
The magnetic field  affects strongly  the NS outer crust \cite{Chamel11,Potekhin13,Chamel15,Stein16}, in particular, the neutron drip density, its composition and the properties of the nuclei present in the outer crust.

 In Refs. \cite{Fang16,Fang17,Chen17,Fang17a,Ferreira21}, it was shown  from the calculation of the dynamical or thermodynamical spinodals that a complex inner crust could exist with several disconnected non-homogeneous regions. Considering a fixed proton fraction, it was shown that disconnected regions with pasta phases could exist at densities above the one associated to the $B=0$ crust-core transition \cite{Pais2021}. However, it was still necessary to confirm these results if $\beta$-equilibrium would be imposed.

 Avancini {\it et al.} have studied the inner-crust pasta phases in the field-free case \cite{cini,avancini02},  at zero and finite temperatures using the Thomas-Fermi (TF) and coexistence phases (CP) methods within a relativistic mean-field (RMF) description of nuclear matter. Both $\beta$-equilibrium stellar matter and matter with  a fixed proton fraction have been considered.  Later, the effect of the magnetic fields on the inner-crust was also studied by some authors \cite{lima,nandi,rabhi,bao}. In Ref. \cite{lima}, quantities such as nuclear size, surface tension and the transition between pasta configurations, were studied using the TF approximation with a fixed proton fraction of $Y_p$=0.1, 0.3, and considering the RMF NL3 model. Recently, Bao et al \cite{bao} investigated the effects of strong magnetic fields on the pasta properties and crust-core(CC) transition, using the TF approximation and two RMF models, TM1 and IUFSU, imposing the condition of $\beta$-equilibrium. Some features, such as an increase of the proton fraction or the decrease of the binding energy per nucleon, due to the magnetic field, were discussed \cite{bao}.
 
 In the present work, we will study the innermost part of  the crust in $\beta-$equilibrium, using the CP calculation \cite{cini}, and considering a magnetic field strength $B^{*}$ ranging from $5 \times10^{3}$ to $2 \times10^{4}$, with   $B^{*}=B/B^e_c$,  $B^e_c$ being  the critical field at which the electron cyclotron energy is equal to the electron mass,  $B^e_c=4.414\times10^{13}$ G.
In particular, we are interested in confirming whether the disconnected non-homogeneous regions exist above the $B=0$ crust-core transition, and in understanding the properties of clusters inside these regions.
We will only study the effect on the inner crust and not in the outer crust.

The present paper is organized as follows: in section II the methods and the formalism are given, in section III we show  our results and discussion, and we draw some conclusions in section IV.

\section{FORMALISM}	
We describe nuclear matter at the NS inner crust within  a relativistic mean field 
approach, in which the nucleons interact via the exchange of mesons. The
exchanged mesons are the isoscalar-scalar and vector mesons ($\sigma$ and
$\omega$, respectively) and the isovector meson ($\rho$). We consider a system of protons and neutrons
with mass $M$ interacting with and through an isoscalar-scalar field $\phi$ with mass $m_{s}$, an isoscalar-vector 
field $V^{\mu}$ with mass $m_{v}$, an isovector-vector field ${\bf b}^{\mu}$ with mass $m_{\rho}$. We also include a system 
of electrons with mass $m_{e}$ to obtain a charge neutral system. Protons and electrons interact through the electromagnetic 
field $A^{\mu}$. The onset of muons occurs above the crust-core transition and, therefore, they have not been included in the present study. The Lagrangian density reads:

\begin{equation}\label{eq3.2}
	\mathcal{L} = \sum_{i=p,n} \mathcal{L}_{i} + \mathcal{L}_{e} + \mathcal{L}_{\sigma} + \mathcal{L}_{\omega} + \mathcal{L}_{\rho} + \mathcal{L}_{\gamma} \ \ ,  
\end{equation}
\noindent
where the nucleon Lagrangian reads
\begin{equation}
	\mathcal L_i  =  \bar{\psi}_{i} \left[\gamma_{\mu} i D^{\mu} - M^{*}-\frac{1}{2}\mu_N\kappa_b\sigma_{\mu \nu} F^{\mu \nu} \right] \psi_{i} ,
\end{equation}
with
\begin{eqnarray}
iD^\mu = i\partial^\mu - g_{v}V^{\mu} - \frac{g_{\rho}}{2} {\vec \tau} \cdot {\bf b}^{\mu} - e \frac{1+\tau_{3}}{2} A^{\mu}  , 
\end{eqnarray}
and
\begin{eqnarray}
	& M^* = M - g_{s}\phi ,
\end{eqnarray}
the nucleon effective mass.
The electron Lagrangian is given by
\begin{equation}
	{\cal L}_{e}  =  \bar{\psi}_{e} [\gamma_{\mu}(i\partial^{\mu} + eA^{\mu}) - m_{e} ]\psi_{e} \ , \\
\end{equation}
and the meson Lagrangian densities are
\begin{eqnarray}
	{\cal L}_{\sigma} & = &\frac{1}{2} \left(\partial_{\mu}\phi \partial^{\mu} \phi - m_{s}^2 \phi^2 - \frac{1}{3} \kappa \phi^3 - \frac{1}{12} \lambda \phi^4 \right) \ , \\
	{\cal L}_{\omega} & = & -\frac{1}{4} \Omega_{\mu\nu} \Omega^{\mu\nu} + \frac{1}{2}m_{v}^2 V_{\mu}V^{\mu} + \frac{\xi}{4!} g_{v}^4 (V_{\mu}V^{\mu})^2  \ \  \\
	{\cal L}_{\rho} & = & -\frac{1}{4} {\bf B}_{\mu\nu} \cdot {\bf B}^{\mu\nu} + \frac{1}{2}m_{\rho}^2 {\bf b}_{\mu} \cdot {\bf b}^{\mu} \ , \\
	{\cal L}_{\gamma} & = & -\frac{1}{4} F_{\mu\nu} F^{\mu\nu} \ ,
\end{eqnarray}
where the tensors are given by
\begin{eqnarray}\label{tensores}
	\Omega_{\mu\nu} & = & \partial_{\mu}V_{\nu} - \partial_{\nu}V_{\mu}  \ , \\
	{\bf B}_{\mu\nu} & = & \partial_{\mu}{\bf b}_{\nu}   -  \partial_{\nu}{\bf b}_{\mu}  - g_{\rho}({\bf b}_{\mu} \times {\bf b}_{\nu}) \ , \\
	F_{\mu\nu} & = & \partial_{\mu} A_{\nu} - \partial_{\nu}A_{\mu} \ \ .
\end{eqnarray}
 The parameters of the model are: the nucleon mass $M$, three coupling
constants $g_{s}$, $g_{v}$, and $g_{\rho}$, of the mesons to the
nucleons, the electrons mass $m_{e}$, the masses of the mesons
$m_{s}$, $m_{v}$, $m_{\rho}$, and the self-interacting coupling constants
$\kappa$, $\lambda$, and $\xi$.
The electromagnetic coupling constant is given by $e=\sqrt{4\pi/137}$, and
$\tau_{3}=\pm 1$ is the
isospin projection for protons ($+1$) and neutrons ($-1$). The nucleon anomalous magnetic moments (AMM) are introduced via the coupling of the baryons to the electromagnetic
field tensor $F_{\mu\nu}$, with $\sigma_{\mu\nu}=\frac{i}{2}\left[\gamma_{\mu},
\gamma_{\nu}\right] $, and strength $\kappa_{b}$, with
$\kappa_{n}=-1.91315$ for the neutron, and $\kappa_{p}=1.79285$ for the
proton. $\mu_N$ is the nuclear magneton. The contribution of the  anomalous magnetic moment of the
electrons is negligible \cite{duncan00}, hence it will not be considered.

In this work, we use two RMF models, NL3  \cite{nl3} and NL3$\omega\rho$  \cite{hor01,pais16V}. 
 The last model contains an additional nonlinear term ${\cal L}_{\omega\rho}$, that mixes the $\omega$ and $\rho$ mesons, allowing to soften the density dependence of the symmetry energy above saturation density. This term is given by
\begin{eqnarray}
	\mathcal{L}_{\omega \rho } &=& \Lambda_v g_v^2 g_\rho^2 V_{\mu }V^{\mu }
	\mathbf{b}_{\mu }\cdot \mathbf{b}^{\mu }.
\end{eqnarray}
These two models belong to the same family, meaning that they have the same isovector saturation properties, i.e. the binding
energy, $E_b=-16.2$ MeV, the saturation density, $\rho_0=0.148$ fm$^{-3}$, and the
incompressibility, $K= 272$ MeV. 
The symmetry energy, $J$, and its slope, $L$, differ, and they are equal to $J=31.7$ (37.4)  MeV, and $L= 55.5$ (118.9) MeV  for NL3$\omega\rho$ (NL3). The  NL3$\omega\rho$ model satisfies the constraints imposed by microscopic calculations of neutron matter \cite{neutron} although NL3 does not.  Both predict stars with masses above 2$M_\odot$, even when hyperonic degrees of freedom are considered \cite{fortin16}.
The choice of two models with the same isoscalar properties will allow us to study the effect of the density dependence of the symmetry energy on the non-homogenous matter. The symmetry energy  is the same at 0.1 fm$^{-3}$ for both models, but, at sub-saturation densities, the two models show a quite different density dependence, see Fig. 1 of ref. \cite{grill}. Let us stress that although the high density  behavior of NL3 is not able to describe the tidal deformability obtained from  GW170817 or the NS radii determined from NICER observations,  the NL3 parametrization was fitted to a large number of nuclear properties \cite{nl3} and should be adequate to study systems that have a density similar to the ones found in nuclei, i.e. sub-saturation densities, as the ones occurring in the inner crust.

The field equations of motion follow from the Euler-Lagrange equations. From the Lagrangian density in Eq.~(\ref{eq3.2}), we obtain the following meson field equations in the mean-field approximation 
\bea
m^{2}_s\left\langle \phi \right\rangle &=&g_s\left(\rho^{s}_{p}+\rho^{s}_{n}\right) 
=g_s\rho^{s},\label{mes1} \\
m^{2}_{v} \left\langle V^{0}\right\rangle  &=& g_v\left(\rho^v_{p}+\rho^v_{n}\right)= 
g_v\rho_{b},\label{mes2} \\
m^{2}_{\rho} \left\langle b^{0}\right\rangle  &=& \frac{1}{2}g_{\rho}\left(\rho^v_{p}-\rho^v_{n}\right) =\frac{1}
{2}g_{\rho}\rho_{3},\label{mes3}
\eea
and the Dirac equations for nucleons and electrons are given by 
\bea
(i\gamma_{\mu}\partial^{\mu}-q_{b}\gamma_{\mu}A^{\mu}- M^* 
-g_v\gamma_{0}V^{0} \cr
-\frac{1}{2}g_{\rho}\tau_{3 b}\gamma_{0}b^{0}-
\frac{1}{2}\mu_{N}\kappa_{b}\sigma_{\mu \nu} F^{\mu \nu}) \Psi_{b}&=&0, \label{MFbary}\\
\left(i\gamma_{\mu}\partial^{\mu}-q_{e}\gamma_{\mu}A^{\mu}-m_{e} \right) \psi_{e}&=&0 \, . \label{MFlep}
\eea
$\rho^{s}$ is the total scalar number density and $\rho_b$ is the total baryonic density. In charge-neutral, $\beta$ equilibrium matter, the conditions 
\beq
\rho^{v}_{p}=\rho^{v}_{e}, \label{neutra}
\eeq
\beq
\mu_n=\mu_p+\mu_e, \label{beta1}
\eeq
should be satisfied.
We solve the coupled Eqs.~(\ref{mes1})-(\ref{neutra}) self-consistently at a given baryon density in the presence of strong magnetic fields. The energy 
density of neutron star matter is given by 
\beq
\varepsilon=\sum_{b=p,n} \varepsilon_{b}+\varepsilon_{e}+\frac{1}
{2}m^{2}_{s}\phi^{2}+\frac{1}{2}m^{2}_{v}V^{2}_{0}+\frac{1}{2}m^{2}_{\rho}b^{2}_{0},
\eeq 
where the energy densities of nucleons and electrons have the following forms
\bea
\varepsilon_{p}&=&\frac{q_{p}B}{4\pi^ {2}}\sum_{\nu=0}^{\nu_{\mbox{\small max}}}\sum_{s}\bigg[k^{p}_{F,\nu,s}E^{p}_{F}
+\Big(\sqrt{m^{* 2}_{p}+2\nu q_{p}B}\Big.\notag\\&&\Big.-s\mu_{N}\kappa_{p}B\Big) ^{2}
\ln\left|\frac{k^{p}_{F,\nu,s}+E^{p}_{F}}{\sqrt{m^{* 2}_{p}+2\nu q_{p}B}-s\mu_{N}\kappa_{p}B} \right|\bigg] , \cr
\varepsilon_{n}&=&\frac{1}{4\pi^ {2}}\sum_{s}\bigg[\frac{1}{2}k^{n}_{F, s}E^{n 3}_{F}-\frac{2}
{3}s\mu_{N}\kappa_{n} B E^{n 3}_{F}\Bigg(\arcsin\left(\frac{\bar{m}_{n}}{E^{n}_{F}} \right)\Bigg.\notag\\&&\Bigg.-\frac{\pi}{2}\Bigg)-\left(\frac{1}{3}s\mu_{N}\kappa_{n} B +\frac{1}{4}\bar{m}_{n}\right) \Bigg(\bar{m}_{n}k^{n}_{F, s}E^{n}_{F}\Bigg.\notag\\&&\Bigg.+\bar{m}^{3}_{n}\ln\left|\frac{k^{n}_{F,s}+E^{n}_{F}}{\bar{m}_{n}}
\right|\Bigg) \bigg], \cr
\varepsilon_{e}&=&\frac{|q_{e}|B}{4\pi^ {2}}\sum_{\nu=0}^{\nu_{\mbox{\small max}}}\sum_{s}\bigg[k^{e}_{F,\nu,s}E^{e}_{F}
+\left(m^{2}_{e}+2\nu |q_{e}|B\right)\cr&&\ln\left|\frac{k^{e}_{F,\nu,s}+E^{e}_{F}}{\sqrt{m^{2}_{e}+2\nu |q_{e}| B}} \right|\bigg]. 
\eea

The expressions of the scalar and vector densities for protons and neutrons are given by
\bea
\rho^{s}_{p}&=&\frac{q_{p}Bm^{*}_{p}}{2\pi^{2}}\sum_{\nu=0}^{\nu_{\mbox{\small max}}}\sum_{s}\frac{\sqrt{m^{* 2}_{p}+2\nu 
		q_{p}B}-s\mu_{N}\kappa_{p}B}{\sqrt{m^{* 2}_{p}+2\nu q_{p}B}}\cr
	& &\times\ln\left|\frac{k^{p}_{F,\nu,s}+E^{p}_{F}}
{\sqrt{m^{* 2}_{p}+2\nu q_{p}B}-s\mu_{N}\kappa_{p}B} \right| , \cr
\rho^{s}_{n}&=&\frac{m^{*}_{n}}{4\pi^{2}}\sum_{s} \left[E^ {n}_{F}k^{n}_{F, s}-\bar{m}^{2}_{n}\ln\left|
\frac{k^{n}_{F,s}+E^{n}_{F}}{\bar{m}_{n}} \right|\right],  \cr
\rho^{v}_{p}&=&\frac{q_{p}B}{2\pi^{2}}\sum_{\nu=0}^{\nu_{\mbox{\scriptsize max}}}\sum_{s}k^{p}_{F,\nu,s},  \cr
\rho^{v}_{n}&=&\frac{1}{2\pi^{2}}\sum_{s}\left[ \frac{1}{3}\left(k^{n}_{F, s}\right)^{3}-\frac{1}{2}s\mu_{N}\kappa_{n}B\Bigg(\bar{m}_{n}k^{n}_{F,s} \notag \Bigg.\right.\\&& \left.\Bigg.%
+E^{n2}_{F}\left(\arcsin\left(\frac{\bar{m}_{n}}{E^{n}_{F}}\right)  -\frac{\pi}{2}\right)  \Bigg)\right],
\eea
and the vector densities for electrons are given by
\beq
\rho^{v}_{e}=\frac{|q_{e}|B}{2\pi^{2}}\sum_{\nu=0}^{\nu_{\mbox{\small max}}}\sum_{s}k^{e}_{F,\nu,s},
\eeq
where $k^{p}_{F,\nu,s}$, $ k^{n}_{F,s}$ and $k^{e}_{F,\nu,s}$ are the Fermi momenta, respectively, of protons, neutrons and electrons, which are related to the Fermi energies $E^{p}_{F}$, $E^{n}_{F}$ and  $E^{e}_{F}$ as
\bea
k^{p 2}_{F,\nu,s}&=&E^{p 2}_{F}-\left[\sqrt{m^{* 2}_{p}+2\nu q_{p}B}-s\mu_{N}\kappa_{p}B\right] ^{2}, \cr
k^{n 2}_{F,s}&=&E^{n 2}_{F}-\bar{m}^{2}_{n}, \cr
k^{e 2}_{F,\nu,s}&=&E^{e 2}_{F}-\left(m^{2}_{e}+2\nu |q_{e}| B\right),  \quad \label{eq25}
\eea
with
\beq
\bar{m}_{n}=m^{*}_{n}-s\mu_{N}\kappa_{n}B.\label{barm}
\eeq
The summation in $\nu$ in the above expressions terminates at $\nu_{max}$, the largest value of $\nu$ for which the square of Fermi momenta of the particle is still positive and which corresponds to the closest
integer from below defined by the ratio
$$\nu_{max}=\left[\frac{(E^e_F)^2-m_e^2}{2 |q_e|\, B}\right],\quad \mbox{for electrons}$$
$$\nu_{max}=\left[\frac{(E^p_F+s\,\mu_N\,\kappa_p\,B)^2-{m_p^*}^2}{2 |q_p|\,
	B}\right], \quad \mbox{for protons}.$$
The chemical potentials of baryons and electrons are defined as 
\bea
\mu_{b}&=& E^{b}_{F}+g_{v}V^{0}+\frac{1}{2}g_{\rho}\tau_{3 b}b^{0}, \label{eq27}\\
\mu_{e} &=& E^{e}_{F}=\sqrt{k^{e 2}_{F,\nu,s}+m^{2}_{e}+2\nu |q_{e}| B}.\label{eq28}
\eea

In the coexisting phases method, matter is organized into separated regions of higher and lower density, the higher ones being the pasta phases, and the lower ones, a background nucleon gas, see \cite{avancini02,Pais2021}. The interface between these regions in the present approach is sharp. Finite size effects are taken into account by including a surface and a Coulomb term in the energy density after the minimization of the free energy. 

By minimizing the sum $\varepsilon_{surf}+\varepsilon_{Coul}$ with respect to the size of the droplet/bubble, rod/tube or slab, one gets 
\begin{eqnarray}
	\varepsilon_{surf}&=& 2\varepsilon_{Coul}, \label{surf} 
\end{eqnarray}
with
\begin{eqnarray}
	\varepsilon_{Coul}&=&\frac{2\alpha}{4^{2/3}}\left(e^2\pi\Phi\right)^{1/3}\left[\sigma D(\rho_p^I-\rho_p^{II})\right]^{2/3}, \label{coul}
\end{eqnarray}
where $\alpha=f$ for droplets, rods and slabs and $\alpha=1-f$ for tubes and bubbles, $f$ is the volume fraction of phase $I$, $\sigma$ is the surface energy coefficient and $\Phi$ is given by
\begin{eqnarray}
	\Phi=\left\lbrace\begin{array}{c}
		\left(\frac{2-D \alpha^{1-2/D}}{D-2}+\alpha\right)\frac{1}{D+2},~ \quad D=1,3 \\
		\frac{\alpha-1-\ln \alpha}{D+2}, ~\quad D=2 \quad 
	\end{array} \right. 
\end{eqnarray}
The Gibbs equilibrium conditions are imposed to get the lowest energy state, and, for a temperature $T=T^I=T^{II}$, are written as 
\begin{eqnarray}
	\mu_n^I&=&\mu_n^{II},  \label{cp} \\
	\mu_p^I&=&\mu_p^{II}, \nonumber \\
	P^I&=&P^{II}, \nonumber 
\end{eqnarray}
where $I$ and $II$ label the high- and low-density phases, respectively.

\section{Results and discussion}

 \begin{figure*}[htbp]\centering
	\begin{center}
		\begin{tabular}{lcc}
			\includegraphics[width=1\linewidth,angle=0]{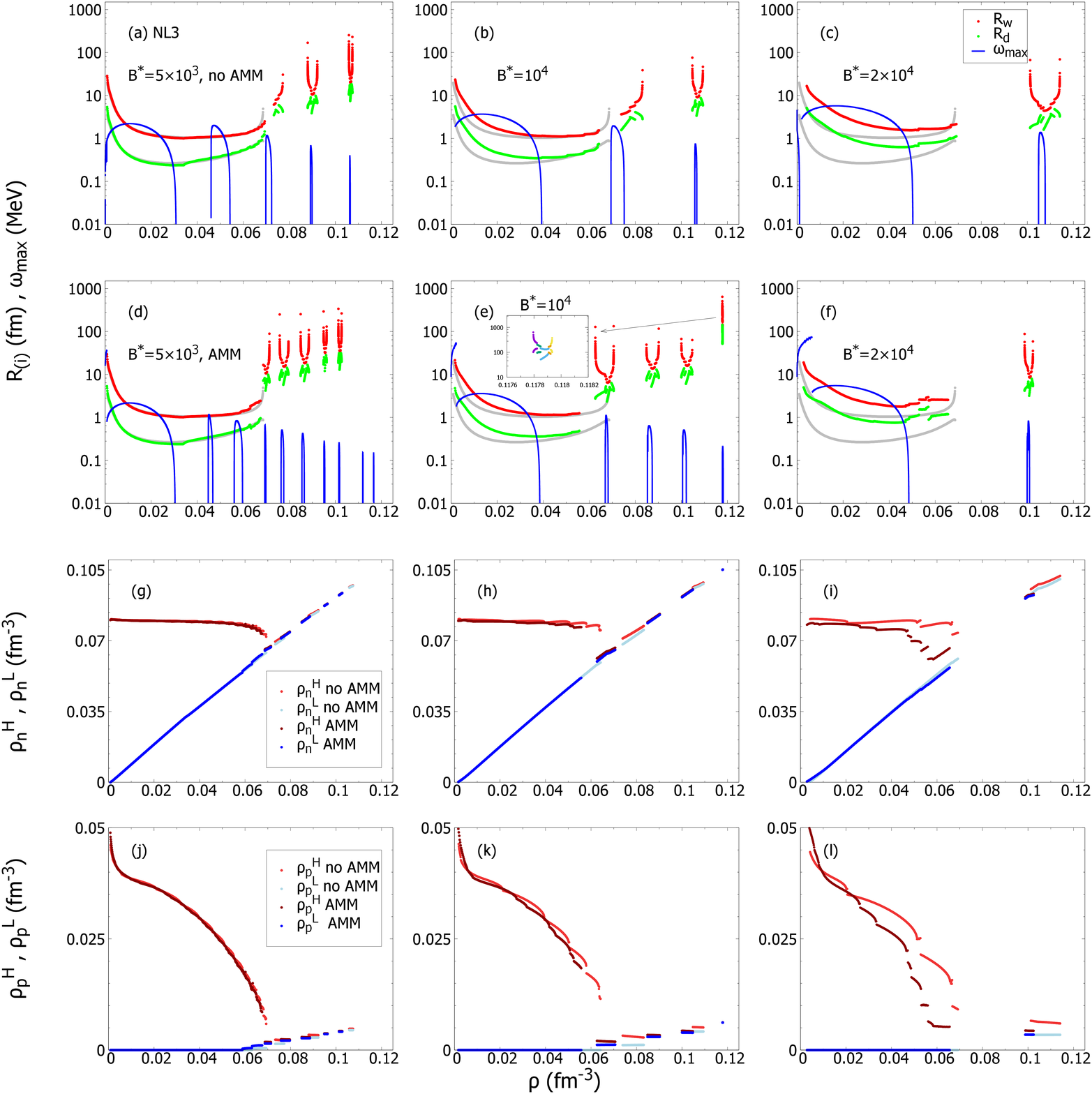}
		\end{tabular}
	\end{center}
	\caption{Radii of WS cell (red) and nucleus (green) for $\beta$-equilibrium matter using  the NL3 parametrization without (first row, (a), (b) and (c)) and with (second row, (d), (e) and (f)) the inclusion of AMM for different magnetic field strengths $B^*$=$5\times10^3$ (left, (a), (d), (g) and (j)), $10^4$ (middle, (b), (e), (h) and (k)), $2\times10^4$ (right, (c), (f), (i) and (l)). The no-field case is also shown with gray points as a reference. Growth rates obtained with a dynamical spinodal calculation in \cite{Fang17a} are plotted with blue lines. The gas (blue), designated by L for low density, and  the cluster (red), designated by H for high density, neutron (third row, (g), (h), (i)) and proton (fourth row, (j), (k), (l)) densities inside the WS cell are also plotted as a function of the denisty without (light colors) and with (dark colors) AMM.
	}
	\label{fig1}
\end{figure*}

\begin{figure*}[htbp]\centering
	\begin{center}
		\begin{tabular}{lcc}
			\includegraphics[width=1\linewidth,angle=0]{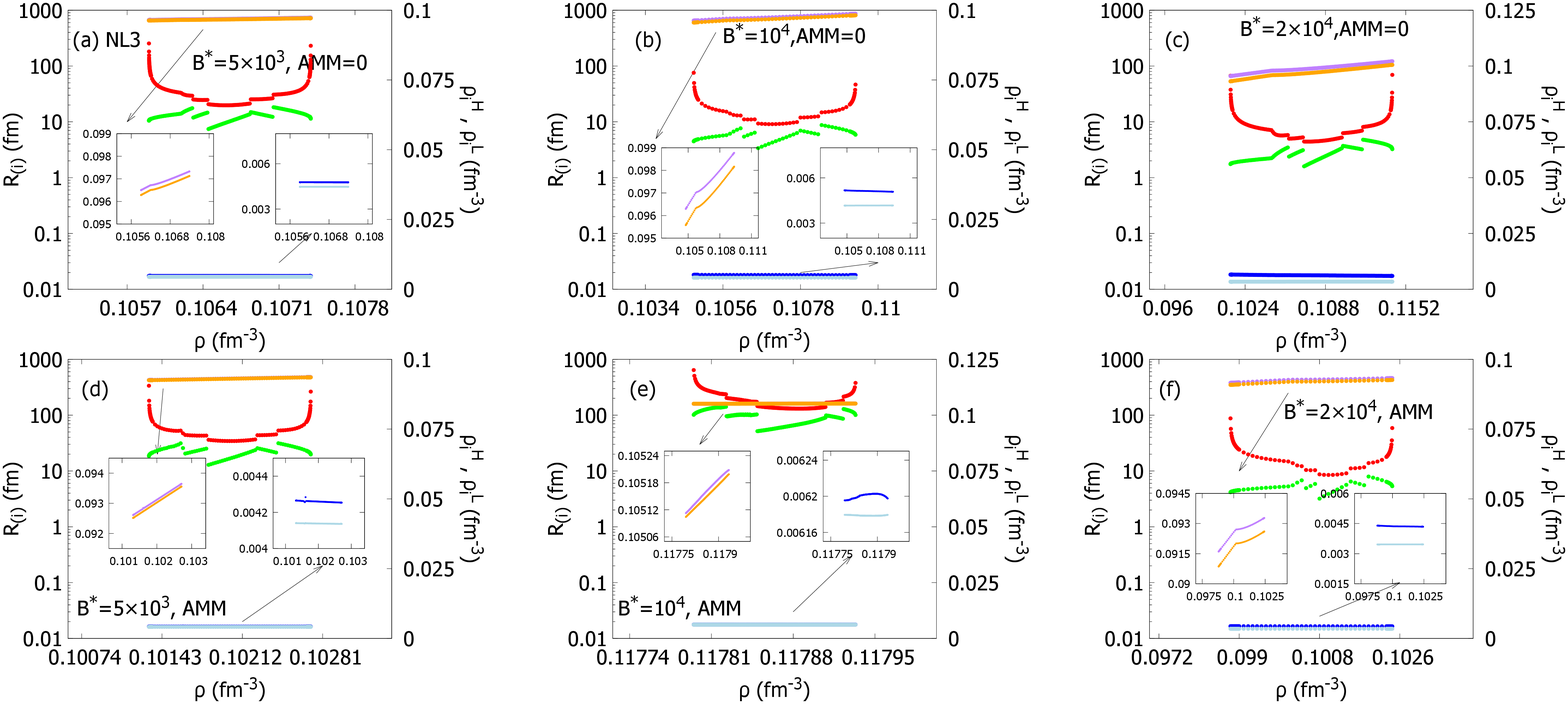}
		\end{tabular}
	\end{center}
	\caption{The pasta phases at the largest-density disconnected region: radii of the WS cell (red) and cluster (green), proton density of high-density phase $\rho_p^H$ (blue) and low-density phase $\rho_p^L$(light-blue), neutron density of the high-density phase $\rho_n^H$(purple), and low-density phase $\rho_n^L$(orange) as a function of the density, for the NL3 parametrization, with (bottom) and without (top) AMM. The insets show that the low and high proton and neutron densities are similar but different. As before, H (red) is for the cluster density and L (blue) for the gas density.
	}
	\label{fig1p}
\end{figure*}

\begin{figure}[htbp]\centering
	\begin{center}
		\begin{tabular}{lcc}
			\includegraphics[width=1\linewidth,angle=0]{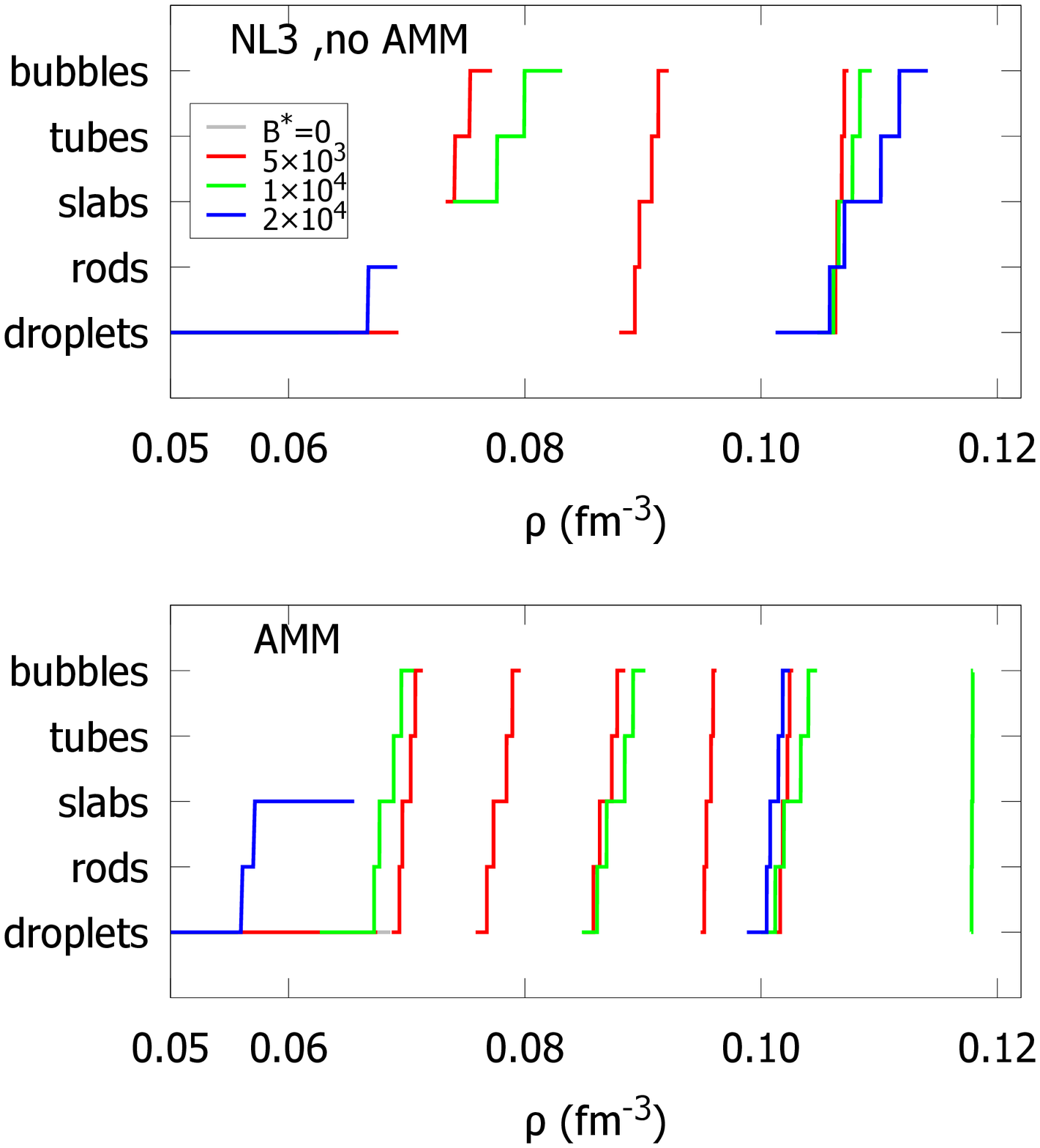}
		\end{tabular}
	\end{center}
	\caption{The evolution of the pasta shapes as a function of the baryonic density for $\beta$-equilibrium  matter and for different magnetic field strenghts: $B^*$=0 (gray), $5\times 10^3$ (red), $10^4$ (green), $2\times10^4$ (blue), for the NL3 model with (bottom) and without (top) AMM.}
	\label{fig2}
\end{figure}

Having as a main objective to understand how the magnetic field affects the structure of the inner crust and the crust-core transition, in the following 
 we calculate  the non-homogeneous matter in the inner crust of NSs in $\beta$-equilibrium under different field strengths, of the order of $\sim$ $10^{17}-10^{18}$G. 
 All calculations are performed  with and without the anomalous magnetic moment (AMM) of the nucleon in order to evaluate under which conditions it is important to include it. 
 
 In Fig.\ref{fig1}, the radii of the WS cells and clusters inside them are shown as a function of density without (top row) and with (middle row) AMM. In the two bottom rows, the neutron and proton densities are given for the cluster and the gas phases, without AMM (light colors) and with AMM (dark colors). All the calculations shown in this Figure are for the NL3 model.
 In the same figure, in the two top rows, the growth rates $\Gamma$ determined within a dynamical spinodal calculation are also plotted with a blue line. The dynamical spinodal formalism has been introduced in \cite{Avancini2005,Providencia2006}, and it is based on the calculation of the small oscillation frequencies of matter obtained by considering small deviations of the fields and distribution functions of neutrons, protons and electrons from equilibrium. The spinodal region is characterized by purely imaginary frequencies $\omega$ and we denominate the growth rates as $\Gamma=|\omega|$.
In Ref.~\cite{Fang16}, the dynamical spinodal approach was applied to magnetized matter and the authors have predicted the occurrence  of  pasta phases at larger densities than the expected ones, i.e. as in the $B=0$ case. In this Figure, we see that other pasta phase regions, disconnected from the low-density spinodal region, appear, contrary to the field-free case. This effect  is caused by the Landau quantization induced by the magnetic field, and, as it was said before, it was predicted in previous works, which pointed out that strong magnetic fields may cause an extension of the NS inner crust \cite{Fang16,Fang17,Chen17,Fang17a}.

\begin{figure*}[htbp]\centering
	\begin{center}
		\begin{tabular}{lcc}
			\includegraphics[width=1\linewidth,angle=0]{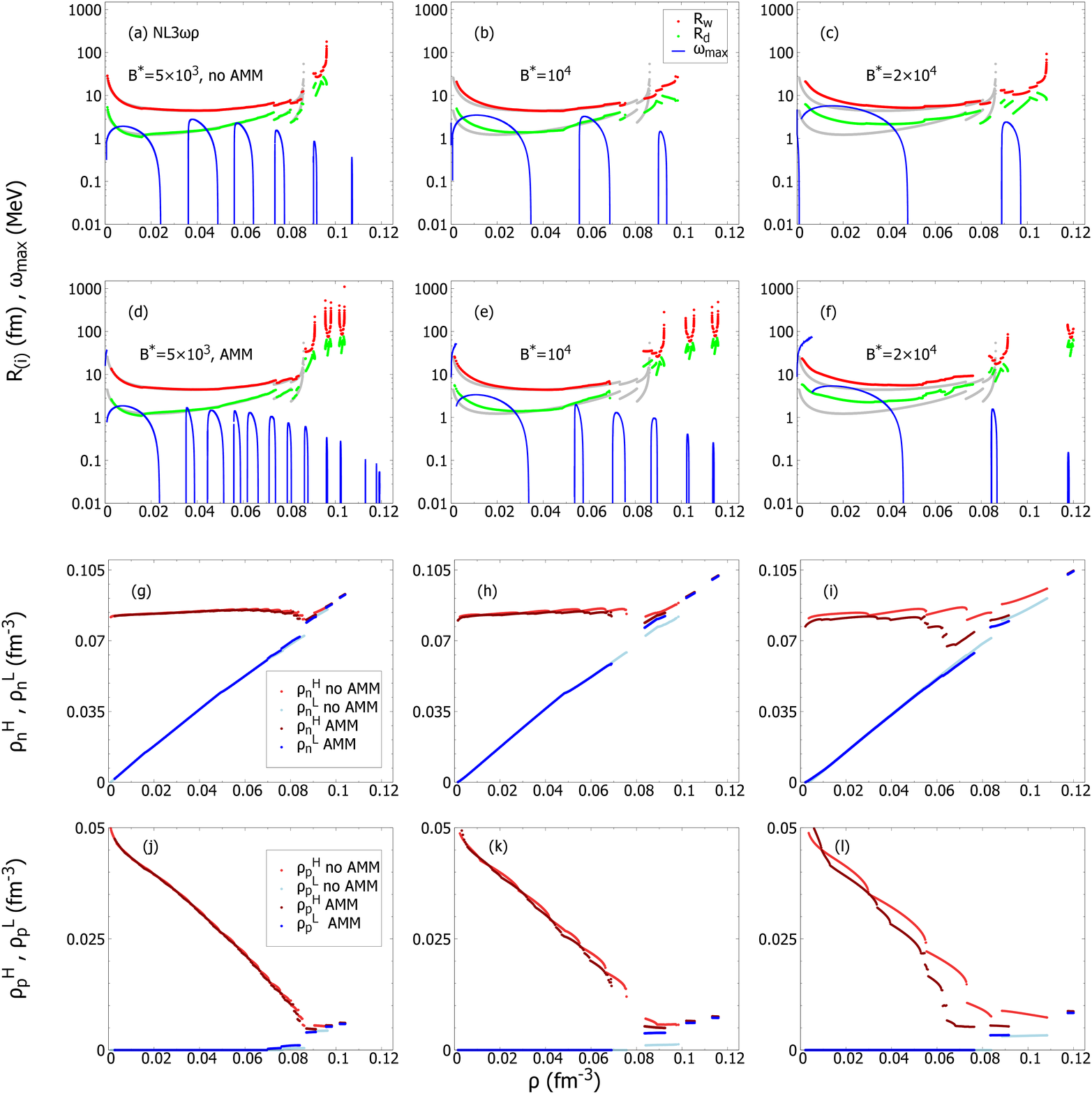}
		\end{tabular}
	\end{center}
	\caption{Radii of the WS cell (red) and nucleus (green) for $\beta$-equilibrium matter using the NL3$\omega\rho $ parametrization without (first row, (a), (b) and (c)) and with (second row, (d), (e) and (f)) the inclusion of AMM for different magnetic field strengths $B^*$=$5\times10^3$ (left, (a), (d), (g) and (j)), $10^4$ (middle, (b), (e), (h) and (k)), $2\times10^4$ (right, (c), (f), (i) and (l)). The no-field case is also shown with gray points as a reference. Growth rates obtained with a dynamical spinodal calculation in \cite{Fang17a} are plotted with blue lines. The gas (L, blue) and the cluster (H, red) neutron third row, (g), (h), (i)) and proton (fourth row, (j), (k), (l)) densities inside the WS cell are also plotted as a function of the density without (light colors) and with (dark colors) AMM.}
	\label{fig3}
\end{figure*}

 In Fig.\ref{fig1}, some interesting features may be identified: (i) as for the "primary" pasta region, namely for a  density $\rho\lesssim 0.06fm^{-3}$ (for NL3$\omega\rho,\rho\lesssim 0.08fm^{-3}$), when the magnetic fields are relatively weak ($B^*$=5$\times10^3$ in the present work), the impact of the magnetic field and AMM are negligible. This is  consistent with the results of Ref. \cite{bao}. In this last study,  Bao et al. found that the effects of the magnetic fields and AMM are negligible when B$\leq10^{17}$G, see also \cite{broderick,rabhi2}; (ii) above the connected low-density instability region, there are several disconnected pasta regions that appear, the stronger the field $B$, the smaller the number of regions and the wider the density range that they cover. As we will see, in each one of these regions, all different geometric configurations may occur; (iii) as for the extended pasta regions (disconnected non-homogeneous matter), it is verified that if AMM is considered, more and narrower disconnected regions occur. This is explained by the spin polarization: including the AMM, the spin polarization degeneracy is removed; (iv) the weaker the fields, the closer the disconnected pasta regions, and, in general, the narrower they are; (v) for a given magnetic field strength $B^*$, the higher the baryon density, the narrower the individual pasta regions. A very narrow pasta is visible at the density of $\sim 0.117$ fm$^{-3}$ in Fig. \ref{fig1}(e) (see the magnification  in the inset). We should point out that for much weaker fields, like $B^*$=$10^3$ and 3$\times 10^3$, more structures like these would occur and be closer together, which is in agreement with the above conclusions. We recall that $B^*$=$10^3$ ($5\times 10^3$) corresponds to a field $B=4.4\times 10^{16}$~G ($B=2.2\times 10^{17}$~G).
 
\begin{figure*}[htbp]\centering
	\begin{center}
		\begin{tabular}{lcc}
			\includegraphics[width=1\linewidth,angle=0]{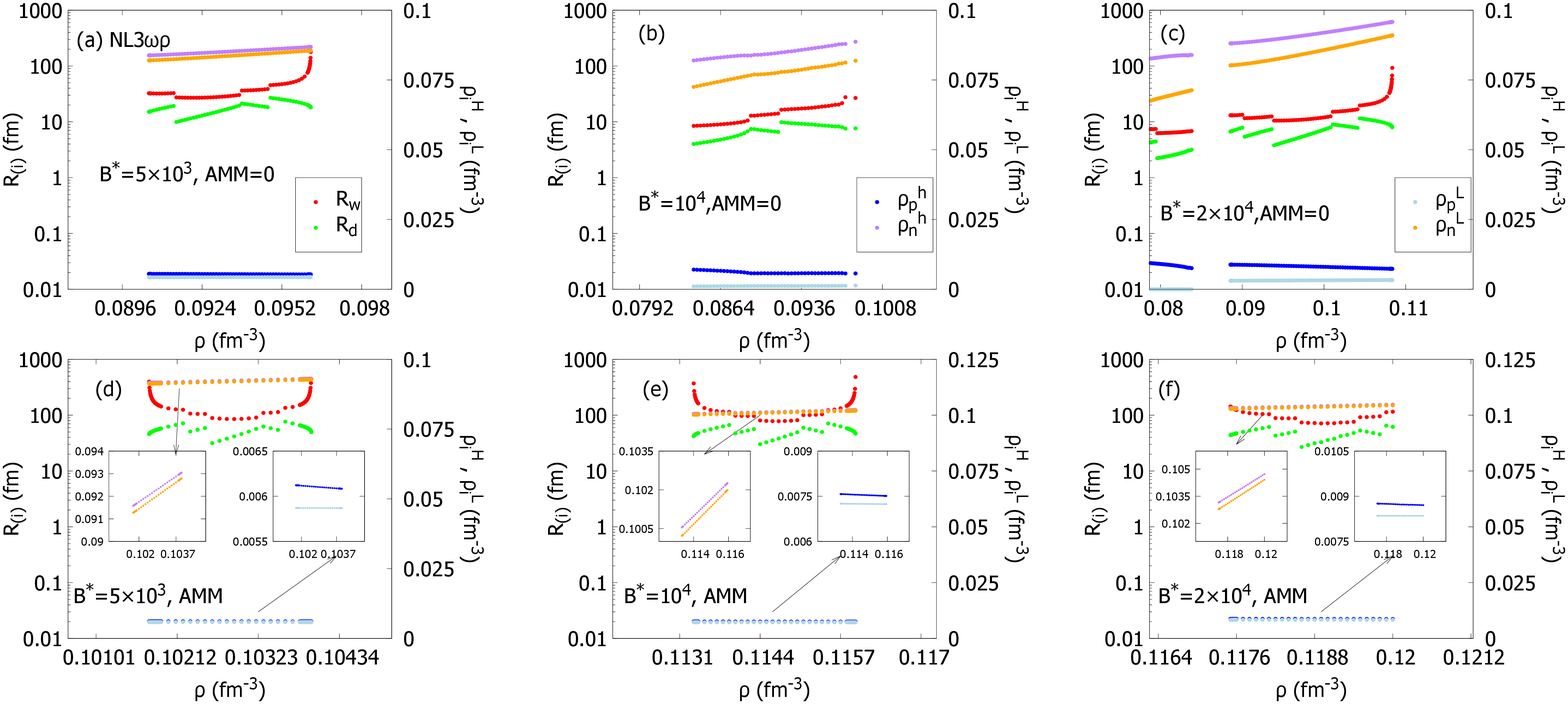}
		\end{tabular}
	\end{center}
	\caption{The pasta phases at the largest-density disconnected region: radii of the WS cell (red) and cluster (green), proton density of high-density phase $\rho_p^H$ (blue) and low-density phase $\rho_p^L$(light-blue), neutron density of the high-density phase $\rho_n^H$(purple), and low-density phase $\rho_n^L$(orange) as a function of the density, for the NL3$\omega\rho$ parametrization, with (bottom) and without (top) AMM. The insets show that the low and high proton and neutron densities are similar but different. }
	\label{fig3p}
\end{figure*}

\begin{figure}[ht]\centering
	\begin{center}
		\begin{tabular}{lcc}
			\includegraphics[width=1\linewidth,angle=0]{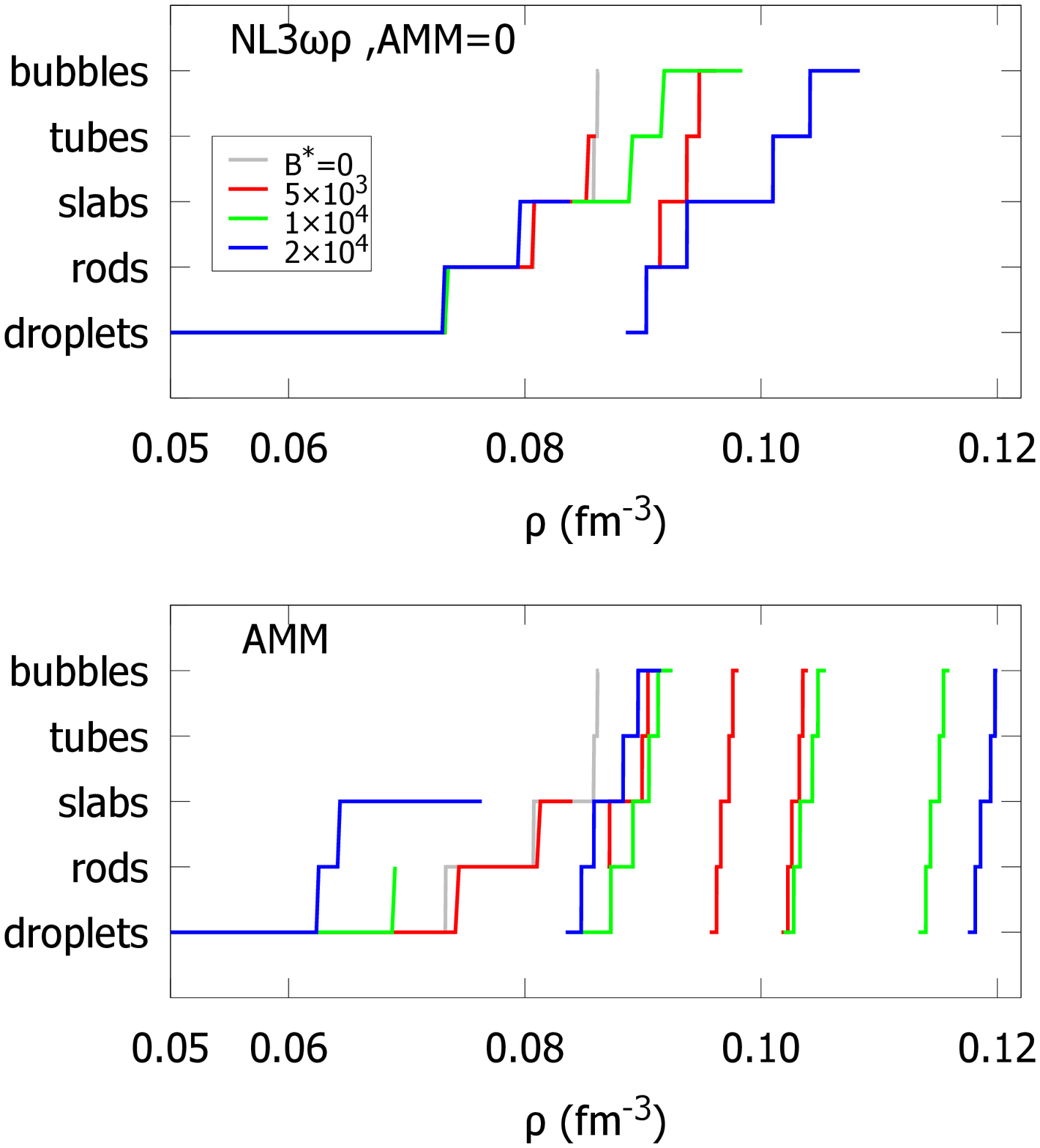}
		\end{tabular}
	\end{center}
	\caption{The evolution of the pasta shapes as a function of the baryonic density for $\beta$-equilibrium  matter and for different magnetic field strenghts: $B^*$=0 (gray), $5\times 10^3$ (red), $10^4$ (green), $2\times10^4$ (blue), for the NL3$\omega\rho$ model with (bottom) and without (top) AMM.}
    \label{fig4}
\end{figure}

\begin{figure}[htbp]\centering
	\begin{center}
		\begin{tabular}{lcc}
			\includegraphics[width=1\linewidth,angle=0]{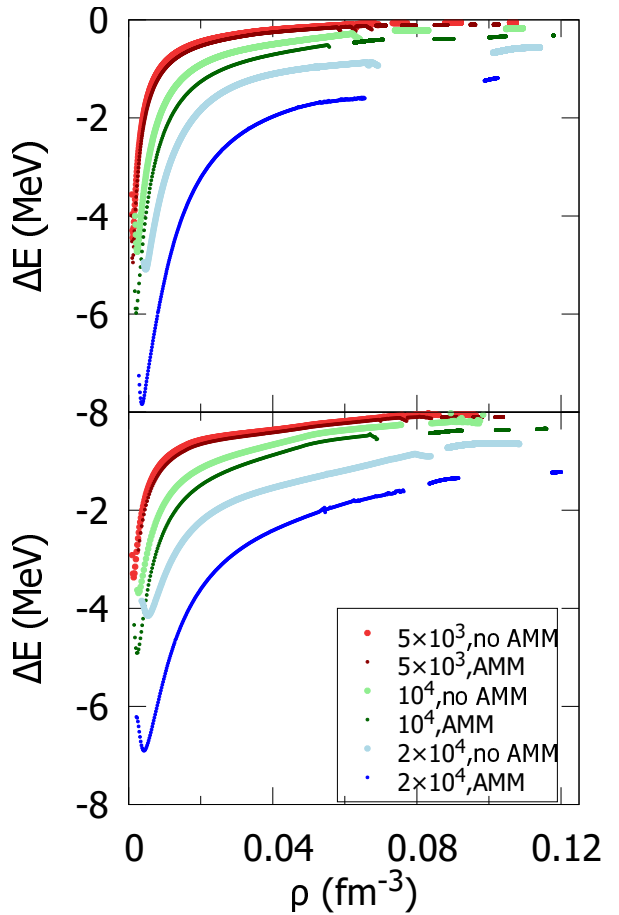}
		\end{tabular}
	\end{center}
	\caption{The difference between the finite-$B$ binding energy per nucleon  of the pasta phases ($E_p$) and the zero-$B$ binding energy per nucleon of homogeneous matter ($E_h$), $\Delta E= E_p(B)-E_h(B=0)$, as a function of the baryonic density for $\beta$-equilibrium  matter using the NL3 (top) and NL3$\omega\rho$ (bottom) parametrizations, considering different magnetic field strengths: $5\times10^3$ (red), $10^4$ (green), and $2\times10^4$ (blue). The results consider calculations with (dark colors) and without (ligh colors) AMM.}
	\label{fig5p}
\end{figure}

\begin{figure}[htbp]\centering
	\begin{center}
		\begin{tabular}{lcc}
			\includegraphics[width=1\linewidth,angle=0]{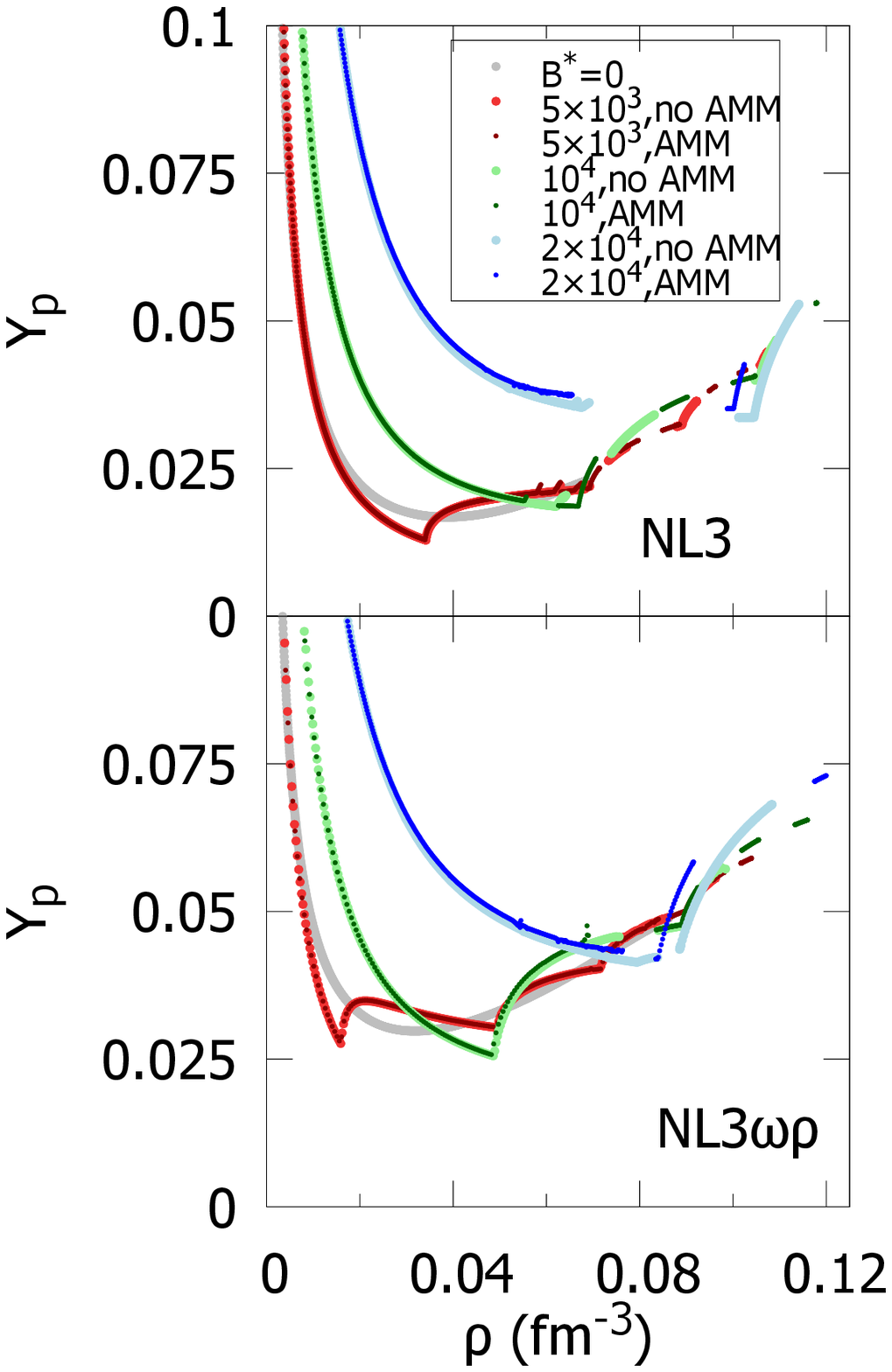}
		\end{tabular}
	\end{center}
	\caption{The proton fraction as a function of the baryonic density for $\beta$-equilibrium  matter using the NL3 (top) and NL3$\omega\rho$ (bottom) parametrizations, considering different magnetic field strengths: $B^*$=0 (gray), $5\times10^3$ (red), $10^4$ (green), and $2\times10^4$ (blue). The results consider calculations with (dark colors) and without (light colors) AMM.}
	\label{fig7}
\end{figure}
 
 In the two bottom rows, the neutron and proton densities inside the WS cells are plotted. The behavior of the neutron gas and cluster densities in the first spinodal region follow the behavior discussed in \cite{grill} and \cite{bao}: the cluster density remains almost flat  while the gas density increases almost linearly. 
 
 Above this first spinodal region the disconnected regions appear which present some special properties: the neutron cluster density increases and the gas and cluster densities differ only slightly, but this difference increases for stronger fields and it is larger if AMM is set to zero. In the bottom row, the proton gas and cluster densities are given. In the first spinodal region, the proton cluster density decreases with density, as in \cite{grill,bao}. 
 Above this region, the proton cluster density remains approximately constant. The presence of the magnetic field has a particular effect on the proton gas density: it is non-zero above $\approx 0.05$ fm$^{-3}$. Just as it was discussed before for the neutrons,  also the proton gas and  the proton cluster densities are almost equal  in these disconnected regions, in particular, if AMM is finite.

 The finite proton density in the background gas will certainly affect matter properties such as the electrical conductivity.

 The structure of the disconnected regions for each magnetic field with and without AMM
 is better understood looking at Fig.~\ref{fig2}. Matter in these disconnected  regions shows the normal configuration transition order, i.e. "droplet-rod-slab-tube-bubble", whereas sometimes reversed-order pasta may arise, namely, in the order of "bubble-tube-slab-rod-droplet", as seen in the Fig. \ref{fig2} of Ref. \cite{Pais2021}. However, in this last work, a fixed proton fraction was considered while in our case the proton fraction increases with density. As discussed in Ref.~\cite{bao}, although for $B=0$, NL3 only shows the droplet configuration,  for a finite magnetic field, all geometries occur in the first spinodal region. Besides, also inside the narrow disconnected spinodal regions generally all the geometries occur for the magnetic fields considered, independently of having or not zero AMM.  This is better seen in Fig.~\ref{fig1p} where the highest-density  disconnected spinodal region is plotted.  We also include the proton and the neutron gas and cluster densities:  notice that, for each species, the cluster and gas densities are very similar, even if the AMM is included (the insets show the difference). Another feature observed is that the transition density to homogeneous matter at the end of the first spinodal region occurs at smaller densities for larger fields: the gray lines in Fig.~\ref{fig1} extends to larger densities than the colored green and red lines. This same conclusion had been drawn in Ref.~\cite{bao}, where within the Thomas-Fermi approach only this spinodal region has been calculated.
 
 The same  quantities plotted in Figs.~\ref{fig1}, \ref{fig1p} and  \ref{fig2} for the NL3 model are also shown in Figs.~\ref{fig3}, \ref{fig3p}  and \ref{fig4}  for the  NL3$\omega\rho$ parametrization. Most of the conclusions mentioned above also hold here. We notice, however, that the difference in the symmetry energy properties of both models has noticeable effects: comparing the  density extension of the disconnected pasta regions obtained with NL3$\omega\rho$ and  NL3 parametrizations, we see that NL3 presents a larger total extension, possibly associated with the larger symmetry energy slope $L$. This effect had already been discussed in Refs.~\cite{Fang16,Fang17,Fang17a}, but in these works, the proton fraction was held constant. Now we confirm the same effect in $\beta$-equilibrium matter. In summary, although for $B=0$, NL3 and  NL3$\omega\rho$ have quite different crust-core transitions, $\approx$ 0.06 and 0.08 fm$^{-3}$, respectively, for the magnetic field intensities shown in Figs.~\ref{fig1} and \ref{fig3}, the extension is similar, and lies in the range $0.1 - 0.12$ fm$^{-3}$ for both models. It should be pointed out that although at saturation NL3 has a symmetry energy larger than the one of NL3$\omega\rho$, respectively 37.4 and 31.7~MeV, in the range of $0.1 - 0.12$~fm$^{-3}$ their symmetry energy is similar. 
 
For the scenarios considered with or without AMM, but relatively weak fields, the end density of droplets within the NL3$\omega\rho$ model  can go almost up to around 0.07 fm$^{-3}$. This behaviour was also obtained in Refs.~\cite{grill2014,bao2}, in two calculations that did not take into account the magnetic fields. The extended pasta regions above that density are difficult to obtain  for the field intensity $B^*=5\times10^3$.  This can be understood, because, as predicted within the growth rates approach, the density range of this clusterized matter is quite small.
The same situation occurs for even weaker fields with strengths as might also occur in the inner crust of magnetars, e.g. $B^*=10^3$ and 3$\times10^3$ (respectively $4\times 10^{16}$ G and $10^{17}$ G). However, our main conclusions may be drawn with the field intensities we consider: for densities above the main spinodal region, which also occurs for zero magnetic field, a finite magnetic field may give rise to disconnected spinodal regions which generally (i) contain all types of geometries in their narrow density range, (ii) with  very similar cluster and gas densities, and (iii) with a nonzero proton density of the gas.

The  binding energy per nucleon  of the pasta phases ($E_p$) with respect to the zero-field 
homogeneous matter binding energy per nucleon ($E_h$), i.e. $\Delta E= E_p(B)-E_h(B=0)$, is plotted  in Fig.~\ref{fig5p} as a function of the baryonic density, using the NL3 (top) and NL3$\omega\rho$ (bottom) models for $5\times10^3$ (red), $10^4$ (green), and $2\times10^4$ (blue). The dark (light) colors correspond to the results with (without) AMM. Both models show a  similar behavior in general: the binding energy is lowered when strong magnetic fields are considered due to the Landau quantization, which causes the softening of the EoS. However, the difference between the binding energies obtained with strong magnetic fields and the field free case become smaller as the  baryonic density increases, because the magnitude of the effect depends on the intensity of the magnetic field with respect to the Fermi momentum of the nucleon: the larger the $B-$field, the larger the effects on the thermodynamic properties of the magnetized matter. For the stronger fields, it is clearly seen that  the inclusion of AMM lowers the energy, particularly at the lowest densities and strongest magnetic fields, being the most prominent effect of AMM. This essentially reflects the presence of the term $s\, \mu_N \kappa_i B$. Notice also that the disconnected non-homogeneous matter show just a modest trend, almost continuous in the continuation of the "primary" pasta, i.e. the binding energy of clusterized matter is just  a smooth line with segments cut off for some densities. In these intervals, matter is homogeneous and the binding energy was plotted only for non-homogeneous matter.

\begin{table*}[htb]
	\caption{
	The pressures $P_n^{i/f}$~(MeV/fm$^3$), and the corresponding baryonic densities $\rho_n^{i/f}$~(fm$^{-3}$) at the $n-$th initial ($i$) pasta-homogeneous phase transition , and at the $n-$th final ($f$) homogeneous-pasta phase discontinuity points, for NL3 parametrization and all the spinodal regions, starting at the upper boundary of the primary spinodal region. The last density/pressure corresponds to the crust-core transition.
}
	\begin{tabular}{cccccccccccccc}\hline
		$B^*$&$AMM $&& $1,f$& $2,i$&$3,f$&$4,i$&$5,f$& $6,i$&$7,f$&$8,i$& $9,f$& $10,i$&$11,f$\\

	 \hline  

		\multirow{4}{*}{$5\times10^3$}&\multirow{2}{*}{No}&$\rho_n^{i,f}$ &0.06930   &0.07330   &0.07722   &0.08800 &0.09217  &0.10590   &0.10739   &...  &...   &...   & ...  \\
		 
		&                         &$P_n^{i/f}$ &0.56126   &0.61657   &0.71852   &1.07596  &1.21360  &1.83551   &1.89440   &~...~   &~...~   &~...~   & ~...~  \\ 
		&\multirow{2}{*}{Yes} &$\rho_n^{i,f}$&0.06750   &0.06872   &0.07136   &0.07584  &0.07965  &0.08493   &0.08849   &0.09490   &0.09622   &0.10132   &0.10271   \\
	
		&                        &$P_n^{i/f}$ &0.49730   &0.52282   &0.57339   &0.68389  &0.79260  &0.96607   &1.09669   &1.32341   &1.37746   &1.60711   &1.67405   \\  
		\hline
		\multirow{4}{*}{$10^4$}&\multirow{2}{*}{No} &$\rho_n^{i,f}$&0.06420   &0.07392   &0.08316   &0.10477  &0.10936  & ...  &...   &...   &...   &...   & ...  \\
		&                         &$P_n^{i/f}$ &0.45504   &0.63916   &0.89136   &1.81560  &1.99741  &...   & ...  & ...  &...   &...   &...   \\ 

		&\multirow{2}{*}{Yes} &$\rho_n^{i,f}$&0.055690   &0.06264   &0.07060   &0.08484  &0.09020  &0.10001   &0.10473   &0.11779   &0.11793   & ...  &...   \\

		&                        &$P_n^{i/f}$ &0.30335   &0.41565   &0.56473   &0.95961  &1.15361  &1.58326   &1.82087   &2.47114   &2.47935   &  ... &...   \\   
	
			\hline
		\multirow{4}{*}{$2\times10^4$}&\multirow{2}{*}{No}&$\rho_n^{i,f}$ &0.06919   &0.10123   &0.11412   & ... & ... &...   &...   &...   &...   &...   &...   \\ 
		 
		&                         &$P_n^{i/f}$ &0.53919   &1.73962   &2.25502   & ... & ... &...   &...   &...   & ...  &...   & ...  \\
		&\multirow{2}{*}{Yes} &$\rho_n^{i,f}$&0.06555   &0.09880   &0.10243   & ... &...  &...   &...   &...   &...   &...   &...   \\ 
		&                       &$P_n^{i/f}$ &0.46980   &1.65124   &1.75850   &...  &...  &...   &...   &...   &...   &...   &...   \\  
			\hline
	\end{tabular}		
	\label{table1}
\end{table*}

\begin{table*}[htb]
	\caption{The pressures $P_n^{i/f}$~(MeV/fm$^3$), and the corresponding baryonic densities $\rho_n^{i/f}$~(fm$^{-3}$) at the $n-$th initial ($i$) pasta-homogeneous phase transition , and at the $n-$th final ($f$) homogeneous-pasta phase discontinuity points, for NL3$\omega\rho$ parametrization and all the spinodal regions, starting at the upper boundary of the primary spinodal region. The last density/pressure corresponds to the crust-core transition. 
	}
	\begin{tabular}{cccccccccccccc}\hline
		$B^*$&$AMM $&& $1,f$& $2,i$&$3,f$&$4,i$&$5,f$& $6,i$&$7,f$&$8,i$& $9,f$& $10,i$&$11,f$\\
		\hline  
		 
		\multirow{4}{*}{$5\times10^3$}&\multirow{2}{*}{No}&$\rho_n^{i,f}$ &0.08516   &0.09531   &0.09622   & ... &...  &...   &...   &...   &...   &...   &...  \\ 
		&                        &$P_n^{i/f}$ &0.96444   &1.10613   &1.25015   & ... & ... & ...  &...   &...   &...   &...   &...   \\
		&\multirow{2}{*}{Yes}&$\rho_n^{i,f}$ &0.08422   &0.08700   &0.09107   &0.09566  &0.09808  &0.10173   &0.10396   &~~...~~&~~...~~ &~~...~~   &~~...~~  \\
		&                     &$P_n^{i/f}$   &0.94087   &1.00968   &1.11587   &1.23839  &1.31097  &1.43031   &1.50674   & ...  &...   &...   & ... \\

		\hline
	
		\multirow{4}{*}{$10^4$}&\multirow{2}{*}{No}&$\rho_n^{i,f}$ &0.07561   &0.08401   & 0.09841  &...  &  ...&...   &...   &...   &...   & ...  & ...  \\ 
		 
		&                       &$P_n^{i/f}$  &0.76215   & 0.96028  &1.29971   &...  &...  & ...  &...   &...   & ...  &...   & ...  \\ 

		&\multirow{2}{*}{Yes} &$\rho_n^{i,f}$&0.06901   &0.08360   &0.09250   &0.10190  &0.10550  &0.11333   &0.11594   &...   &...   &...   & ...  \\
		&                      &$P_n^{i/f}$  &0.62999   &0.95172   &1.16818   &1.44014  &1.56046  &1.85963   &1.96759   &...   & ...  &...   & ...  \\ 
  
		\hline
	
		\multirow{4}{*}{$2\times10^4$}&\multirow{2}{*}{No}&$\rho_n^{i,f}$ &0.08375   &0.09956   &0.10832   &...  &...  &...   &...   &...   & ...  &...   &  ... \\ 
		 
		&                    &$P_n^{i/f}$     &1.05267   &1.16429   &1.63340   &...  &...  &...   &...   & ...  &...   &...   &...   \\ 

		&\multirow{2}{*}{Yes} &$\rho_n^{i,f}$&0.07658   &0.08348   &0.09152   &0.11752  &0.12016  &...   &...   &...   &...   &...   & ...  \\ 
		&                  &$P_n^{i/f}$      &0.86279   &1.06788   &1.18199   &2.05859  &2.16918  &...   &...   &...   &...   &...   &...   \\ 
 
		\hline 
	\end{tabular}		
	\label{table2}
\end{table*}

In Fig. \ref{fig7}, the proton fraction $Y_p$  in the WS cell is plotted against the baryonic density. The proton fraction decreases with increasing density following the $B=0$ proton fraction in the  "primary" pasta region. After attaining a minimum, the proton fraction  starts to increase as shown in Refs.~\cite{grill,bao}. In the presence of a finite magnetic field,  several features are identified: (i) the decrease of the proton fraction at low densities occurs much slower, so that the minimum of $Y_p$ occurs at larger densities; (ii) after the minimum is attained,  the Landau quantization is reflected in the proton fraction, and a fluctuation around the $B=0$ scenario is obtained; (iii) the disconnected spinodal regions have larger proton fractions than the one at the upper boundary of the   "primary" pasta region. This behavior is similar to the one of the homogeneous matter core, above the crust-core transition.
It is evident  that strong magnetic fields increase the proton fraction within the cell, while the difference between AMM and no-AMM cases are barely distinguishable, except at the "extended"  pasta region: slightly larger proton fractions are obtained with AMM. In addition, when the magnetic fields are relatively weak ($B^*$=5$\times10^3\sim 2\times 10^{17}$G), the impact of the magnetic  fields on the proton fraction $Y_p$ is not large, as seen in the Figure, i.e. the red points almost coincide with the gray points until a density of the order of 0.015 fm$^{-3}$. 
  
  In Tables \ref{table1} and \ref{table2}, the  pressures $P_i$~(MeV/fm$^3$) and the corresponding baryonic densities $\rho_i$~(fm$^{-3}$) at the pasta-homogeneous phases or homogeneous-pasta phases transition points are shown, respectively, for  the NL3 and the NL3$\omega\rho$ parametrizations. All the calculated regions have been included in the tables.  
  For a finite magnetic field,  the pressure at the last transition boundary is quite larger than the pressure at the primary spinodal transition to homogeneous matter. 
 The pressure at the crust-core interface $P_t$ is a key parameter to determine the crustal fraction of the moment of inertia \cite{link99,Lat2000,Lat2001,Lat2007,Xu2009}, because the crustal momentum of inertia may be the mechanism that drives glitches  in NSs, as the ones observed in the Vela pulsar \cite{link99}.  It was, however, suggested that this mechanism may be more complicated if the  entrainment effects between the superfluid neutrons and the crust are taken into account \cite{andersson12,chamel12}, and, in this case, the angular momentum reservoir of the crust would not be enough to explain the glitch mechanism. The possible increase of the non-homogeneous matter at the transition to the NS core due to the presence of strong magnetic fields could be an answer to the extra angular momentum reservoir.
Though we may have not computed the crust-core interface precisely, due to convergence problems occurring when the cluster and gas densities are similar, the present study indicates that the pressure at the crust-core transition may be well above the 0.65 MeV/fm$^{3}$ indicated in Ref.~\cite{link99}.

\section{Conclusion}	
In the present paper, we investigate the effects of strong magnetic fields under the $\beta$-equilibrium condition, in the innermost crust of NSs  where  non-homogeneous nuclear matter known as pasta phases may exist. The CP approximation and RMF models NL3 and NL3$\omega\rho$ are employed.  Although the CP approach is non self-consistent, and, in particular, the surface tension has  been considered magnetic field independent, we believe that both qualitative, and even quantitative conclusions, although  probably carrying a large uncertainty, can be drawn. The two models chosen have allowed us to discuss how sensitive is matter with a different symmetry energy behavior to  magnetic field effects.  NL3 and NL3$\omega\rho$, have in common the same isoscalar properties, but have  a very different density dependence of the symmetry energy. NL3 has a smaller symmetry energy below $\rho=0.1$ fm$^{-3}$, where most of the inner crust lies, and this favors smaller proton fractions and stronger magnetic field effects, in particular, a smaller number of Landau levels.  The larger proton fraction predicted by NL3$\omega\rho$ involves a larger number of Landau levels, and a smaller relative extension of the non-homogeneous matter when compared to NL3.

Within a zero-temperature calculation  and the  Wigner-Seitz approximation to describe clusterized matter, we have carried out calculations  with and without the anomalous magnetic moment of  the nucleon. We found, when taking into account strong magnetic fields, that, besides a primary spinodal region occurring at densities as predicted in the $B=0$ calculations \cite{avancini06,alam17,olfa21}, a 
series of disconnected  regions including pasta-like clusters may appear at densities above the first spinodal region, giving rise to an extension of the inner crust of the NS.

Several properties of the non-homogeneous layers have been calculated, including the nuclear size, the pasta shapes, the pasta-homogeneous matter transitions, the binding energy per nucleon, the pressure at the interfaces, and the density-dependence of the proton fraction. 
Most results concerning the first spinodal region are in good agreement with the discussion performed in Ref.~\cite{bao}, where a Thomas-Fermi calculation was used to study the non-homogeneous layers of a magnetized NS. In their study, however,  no disconnected spinodal regions were reported. In previous studies, these disconnected regions had been predicted within a dynamical spinodal, which takes into account both the Coulomb interaction and the finite range of the nuclear force \cite{Fang16,Fang17}. In the present study, the density range of the disconnected spinodal regions, where new pasta phases are found, agree well with the predictions of the dynamical spinodal. Also, in a recent work \cite{Pais2021}, the accordance between the unstable regions obtained within the dynamical approach, and a calculation as the one performed here, was  found. In that calculation, however, a fixed proton fraction was considered, and since the proton fraction increases with the density, and the magnetic field effects are stronger the smaller the proton Fermi momentum, it was not clear if the $\beta$-equilibrium would disfavor the appearance of the high-density disconnected regions. This, however, has proven not to be the case. 

We may summarize the main conclusions of our study as: (i) finite magnetic fields give rise to disconnected spinodal regions at densities above the $B=0$ crust-core transition density; (ii) the extra spinodal regions contain matter in different geometric configurations; (iii) in the WS cells of these spinodal regions, the cluster and gas densities are very close, for both neutrons and protons; (iv) the magnetic field favors the appearance of protons in the background gas of the non-homogeneous matter, even in the first spinodal region;
(v) in the presence of a magnetic field, the pressure at the crust-core transition may be much larger. We also found that the effect of the AMM is weak, except for the largest-field intensity considered, of the order of $9\times 10^{17}$ G.
A self-consistent calculation that describes adequately the range of the nuclear force, as a Thomas-Fermi approach to the description of non-homogeneous  matter, needs to be performed to confirm the present results. The existence of these extra non-homogeneous matter geometries could have a direct effect on the explanation of  the magnetic field evolution inside NSs \cite{lander11,uryu19}.

The present study was performed at zero temperature, however, it is expected that the temperature will wash out the Landau quantization effects which are responsible for the extra spinodal regions. Recently, in Ref.~\cite{Ferreira21}, the effects of temperature under strong magnetic fields within a thermodynamical spinodal calculation was carried out. It is expected that the "extended"  pasta regions  will appear for  temperatures below $10^9$ K for the magnetic fields we consider in our calculation.

\section*{ACKNOWLEDGMENTS}
This work was partly supported by the FCT (Portugal) Projects No. UID/FIS/04564/2020 and POCI-01-0145-FEDER-029912, and by PHAROS COST Action CA16214, and by the Youth Innovations and Talents Project of Shandong Provincial Colleges and Universities (grant No. 201909118). H.P. acknowledges the grant CEECIND/03092/2017 (FCT, Portugal), and J.L. is supported by National Natural Science Foundation of China (Grant No. 11604179), and Shandong Natural Science Foundation (Grant No. ZR2016AQ18).

\end{document}